# Estimation of Nitrogen-to-Iron Abundance Ratios from Low-Resolution Spectra

Changmin Kim[1], Young Sun Lee[2], Timothy C. Beers[3], and Thomas Masseron[4,5]

[1]Department of Astronomy, Space Science and Geology, Chungnam National University, Daejeon 34134, Korea
[2]Department of Astronomy and Space Science, Chungnam National University, Daejeon 34134, Korea
[3]Department of Physics and JINA Center for the Evolution of the Elements, University of Notre Dame, IN 46556, USA
[4]Instituto de Astrofísica de Canarias, E-38205 La Laguna, Tenerife, Spain
[5]Departamento de Astrofísica, Universidad de La Laguna, E-38206 La Laguna, Tenerife, Spain

**Abstract:** We present a method to determine nitrogen abundance ratios with respect to iron ([N/Fe]) from molecular CN-band features observed in low-resolution ($R \sim 2000$) stellar spectra obtained by the Sloan Digital Sky Survey (SDSS) and the Large Sky Area Multi-Object Fiber Spectroscopic Telescope (LAMOST). Various tests are carried out to check the systematic and random errors of our technique, and the impact of signal-to-noise (S/N) ratios of stellar spectra on the determined [N/Fe]. We find that the uncertainty of our derived [N/Fe] is less than 0.3 dex for S/N ratios larger than 10 in the ranges $T_{\rm eff}$ = [4000, 6000] K, log $g$ = [0.0, 3.5], [Fe/H] = [–3.0, 0.0], [C/Fe] = [–1.0, +4.5], and [N/Fe] = [–1.0, +4.5], the parameter space that we are interested in to identify N-enhanced stars in the Galactic halo. A star-by-star comparison with a sample of stars with [N/Fe] estimates available from the Apache Point Observatory Galactic Evolution Experiment (APOGEE) also suggests a similar level of uncertainty in our measured [N/Fe], after removing its systematic error. Based on these results, we conclude that our method is able to reproduce [N/Fe] from low-resolution spectroscopic data, with an uncertainty sufficiently small to discover N-rich stars that presumably originated from disrupted Galactic globular clusters.

## 1. Introduction

The stellar components of the Milky Way (MW) have been recently explored in great detail, based on the large amount of data obtained through spectroscopic surveys such as the Sloan Extension for Galactic Understanding and Exploration (SEGUE; Yanny et al. 2009), the Apache Point Observatory Galactic Evolution Experiment (APOGEE; Majewski et al. 2017), and the Large Sky Area Multi-Object Fiber Spectroscopic Telescope (LAMOST; Cui et al. 2012). In particular, the Galactic stellar halo has been intensively studied, aided by the unprecedented astrometric information from the Gaia mission (Gaia Collaboration et al. 2018). Many substructures, such as Gaia Sausage/Enceladus (GSE; Belokurov et al. 2018; Helmi et al. 2018), and others (Myeong et al. 2018; Koppelman et al. 2019; Myeong et al. 2019; Naidu et al. 2020) have been discovered, implying that many disrupted (or disrupting) dwarf galaxies have contributed to the formation of the present Galactic halo.

The large spectroscopic and astrometric data now available have also triggered the study of the role of the Galactic globular clusters (GCs) in the formation of the Galactic halo. Various studies have reported that member stars in many GCs exhibit enhancements of some light elements such as nitrogen, sodium, and aluminum, and established anti-correlations between carbon and nitrogen, oxygen and sodium, and magnesium and aluminum (Carretta et al. 2009; Lardo et al. 2012; Carretta 2016; Martell et al. 2016; Schiavon et al. 2017; Tang et al. 2020; Horta et al. 2021). This anomalous chemical distribution can be explained by multiple generations of stars formed during the evolution of GCs (e.g., Kayser et al. 2008; Carretta et al. 2009; Smolinski et al. 2011b; Bastian & Lardo 2018).

Among the light elements, the anti-correlation between C and N has emerged as a particularly useful tracer. The spatial distribution and kinematics of N-rich stars have been used to infer how much the GCs have contributed to the buildup of the Galactic halo. Martell et al. (2011) employed the SEGUE spectroscopic data of the molecular CN-band of red giant branch (RGB) stars in the Galactic halo to find that ∼3% of their sample had relatively weak carbon, but strong nitrogen features. They suggested that such stars formed in GCs in the early MW, and were dissolved into the present halo region during a mass-loss process as GCs interacted with the halo. Koch et al. (2019) also derived a similar fraction of 2.6% for the CN-strong stars from the Sloan Digital Sky Survey (SDSS; York et al. 2000) Data Release 14 (DR14; Abolfathi et al. 2018). Because the second-generation stars in GCs exhibit relatively high N abundances, the CN-strong

Corresponding author: Young Sun Lee (youngsun@cnu.ac.kr)

field stars are believed to be accreted from disrupted GCs.

Martell et al. (2011) also claimed, according to their GC formation model, that stars from disrupted GCs should account for at least 17% of the present stellar halo mass of the MW. They reported that the fraction of strong CN-band stars with respect to normal CN-band stars tended to decrease with increasing distance from Galactic center as well, implying that there is more contribution from the GCs to the inner halo than to the outer halo. In a later work, Martell et al. (2016) recognized, by analyzing 253 halo stars observed in APOGEE, that seven stars were enhanced with nitrogen, and that their chemical characteristics are quite different from other stars in the sample, but are similar to those of the stars that belong to the GCs. They suggested from these results that the seven CN-strong stars were born in GCs in the early Galactic formation phase and later accreted into the halo.

This distinct behavior of the N-rich stars was kinematically confirmed by Carollo et al. (2013). They used the stars in Martell & Grebel (2010) and Martell et al. (2011) to contrast the kinematic properties of CN-strong stars with the rest. Their study showed that most of the N-enhanced stars are distributed in the inner halo, and have typical inner-halo kinematics. They even compared their kinematics with the GC kinematics, and found that the N-rich stars have similar dynamics and metallicity ([Fe/H]) to those of GCs. In more recent work, Tang et al. (2020) reported a fraction of 2.5% of CN-strong stars from LAMOST, which is in a good agreement with the estimate of Martell et al. (2011). One interesting finding is that some of the CN-strong stars exhibited strong retrograde motions. They claimed that these objects were accreted from GCs of dwarf galaxies around the MW.

In the past few years, analyses of high-resolution near-infrared data from APOGEE has also been used to identify the presence of N-enhanced stars associated with individual GCs (in particular those in the bulge; Fernandez-Trincado et al. 2019a, 2020a, 2021b), as well as with debris from likely GC disruption in the inner halo (Fernandez-Trincado et al. 2019b, 2021a) and the Magellanic Clouds (Fernandez-Trincado et al. 2020b).

The studies mentioned above provide solid evidence for the GC-origin stars to be one source of the constituents of the Galactic halo population, implying that the GCs must have played some role in assembling the MW, especially in the bulge and the inner halo of the MW.

Even though there has been great advances in this field in recent years, there is still room for more development and progress, as there remain several shortcomings in the previous studies. One main drawback of most of the previous studies (except for the APOGEE high-resolution analyses) is the use of the relative strengths between the CN and CH bands to identify the N-rich objects, rather than using the more intuitive N abundance ratio ([N/Fe]). Although there are some studies (e.g., Lardo et al. 2012, 2016) that attempted to measure the N abundance from low-resolution stellar spectra, they focused on member stars in a globular cluster (NGC 1851) or the Sculptor dwarf spheroidal galaxy rather than the field stars. In addition, as the molecular bands are more subject to the degeneracy of the stellar parameters, a clear choice for the selection of the N-rich stars is not well defined. To overcome this issue, it is necessary to directly measure the N abundance ratio. The most notable feature for the indicator of the N abundance ratio in a low-resolution optical stellar spectrum is the CN-band features around 3880 Å. However, one obstacle to use this band to determine [N/Fe] is that one must know the carbon abundance ratio in advance in order to determine [N/Fe] from the CN features. Owing to this difficulty, most of the previous low-resolution studies for field stars used the line strength of the CN band to classify the N-enhanced stars rather than the [N/Fe] abundance ratio.

Additionally, we need to identify the N-rich stars to the extent possible by measuring the N abundance ratio accurately, in order to facilitate a better understanding of the role of the GC-origin stars in the formation of the Galactic halo. Because the SDSS[6] and LAMOST surveys have produced an unprecedented number of stellar spectra, once we have the capability of measuring [N/Fe] from the stellar spectra obtained by these surveys, we can better characterize the N-rich stars and their association with the assembly of the Galactic halo.

As part of such an effort to study the GC-origin stars, in this paper we develop a method for determining [N/Fe] from low-resolution SDSS and LAMOST stellar spectra. We present our method of determining [N/Fe] in Section 2. In Section 3, we check on the systematic and random errors of our measured [N/Fe] and the impact of the signal-to-noise ratio (S/N) of the spectrum on the estimated [N/Fe]. We validate our method in Section 4 by comparing with stars with high-resolution determinations from APOGEE and the Stellar Abundances for Galactic Archaeology (SAGA; Suda et al. 2008) databases. A summary and conclusions are presented in Section 5.

## 2. METHODOLOGY FOR MEASURING NITROGEN-TO-IRON ABUNDANCE RATIOS

### 2.1. Fundamental Atmospheric Parameters

The goal of this work is to measure the nitrogen-to-iron abundance ratio using the molecular CN-band features around 3883 Å in low-resolution stellar spectra obtained by SDSS and LAMOST. Both surveys have spectral resolution of $R \sim 2000$ and wavelength coverage of 3820 – 9200 Å for SDSS and 3700 – 9000 Å for LAMOST. To estimate [N/Fe] from the CN band in low-resolution stellar spectra, we first need to determine the stellar atmospheric parameters ($T_{\rm eff}$, log $g$, and [Fe/H]). We made use of the most recent version of the SEGUE Stellar Parameter Pipeline (SSPP; Lee et al. 2008a,b, 2011; Allende Prieto et al. 2008; Smolinski et al. 2011) to derive the stellar parameters from the SDSS and LAMOST spectra. As discussed in Lee et al. (2015), this version of the SSPP was upgraded and modified to determine the stellar parameters from LAMOST stellar spectra.

---

[6] Throughout this paper, we collectively refer to all of low-resolution stellar spectra from legacy SDSS, ongoing SDSS, and SEGUE as SDSS data. Although APOGEE is one of the sub-surveys of SDSS, we specifically state its name, as it is a high-resolution spectroscopy survey.

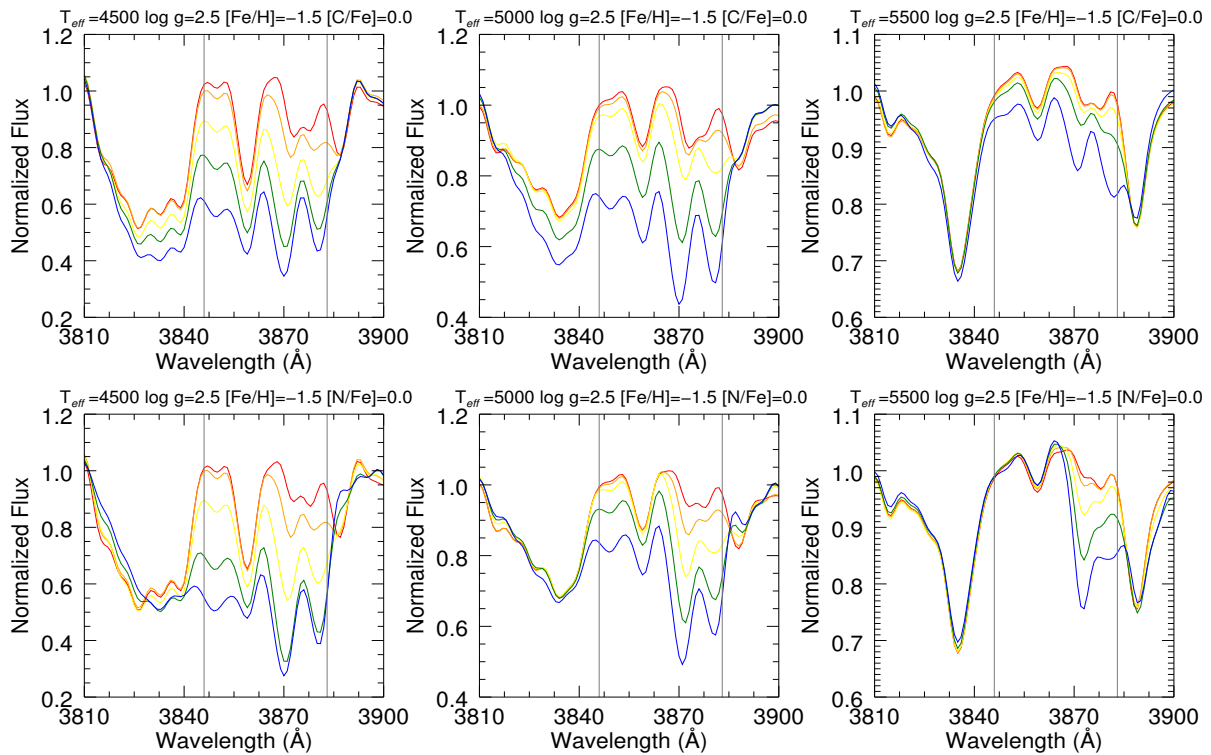

**Figure 1.** Examples of normalized synthetic spectra with different [N/Fe] (top panels) and [C/Fe] (bottom panels) values. The top panels share the same log $g$, [Fe/H], and [C/Fe] of 2.5, –1.5, and 0.0, respectively, but different temperatures, 4500 K, 5000 K, and 5500 K from left to right, while the bottom panels are the same as in the top panels, but for [N/Fe]. The [N/Fe] and [C/Fe] values increase from –0.5 (red) to +1.5 dex (blue) in steps of 0.5 dex. The two vertical lines delineate the CN band (3846 – 3883 Å) used to measure [N/Fe].

Briefly explaining the methodology implemented by the SSPP, it processes the wavelength- and flux-calibrated SDSS and LAMOST stellar spectra, and determines the fundamental stellar parameters for stars in the temperature range 4000 – 10,000 K, using multiple approaches, such as a $\chi^2$ minimization method (Allende Prieto et al. 2006), neural network analysis (Bailer-Jones 2000; Re Fiorentin et al. 2007), and various line-index computations. The introduction of the multiple approaches allows us to use the full spectral range, in order for the SSPP to derive stellar parameters as robust as possible over a wide range in $T_{\rm eff}$, log $g$, [Fe/H], and S/N ratios. The reported typical uncertainty of the SSPP-derived stellar parameters is 180 K, 0.24 dex, and 0.23 dex for $T_{\rm eff}$, log $g$, and [Fe/H], respectively (Smolinski et al. 2011).

### 2.2. Carbon-to-iron Ratios ([C/Fe])

Because the CN bands used to measure [N/Fe] are sensitive to both of the carbon and nitrogen abundances, we need to know in advance not only the stellar atmospheric parameters, but also the carbon-to-iron abundance ratios ([C/Fe]). In order to estimate [C/Fe] from low-resolution SDSS and LAMOST spectra, we adopted the approach used in Lee et al. (2013). They introduced a spectral-matching technique over the molecular CH $G$-band around 4330 Å between a grid of synthetic spectra and an observed spectrum in a $\chi^2$ minimization fashion. The claimed uncertainty in the measured [C/Fe] from SDSS spectra is better than 0.35 dex for stars with S/N > 15 and atmospheric parameters in the ranges $T_{\rm eff}$ = 4400 – 6700 K, log $g$ = 1.0 – 5.0, [Fe/H] = –4.0 – +0.5, and [C/Fe] = –0.25 – +3.5. As this technique is already implemented in the SSPP, once we process the SDSS and LAMOST spectra through the SSPP, the [C/Fe] value is provided along with the stellar parameters.

### 2.3. Determination of Nitrogen-to-iron Abundance Ratios ([N/Fe])

#### 2.3.1. Preparation of a Grid of Synthetic Spectra

Because it is clearly impractical to analyze a large number of stellar spectra one at a time as in high-resolution spectroscopy, we introduce a grid of synthetic spectra to determine the nitrogen-to-iron abundance ratio from SDSS and LAMOST spectra in a fast, efficient manner, following a similar spectral-matching technique as described in Lee et al. (2013). For this approach, we require model atmospheres and their corresponding synthetic spectra. Thus, we first generated model atmospheres with different levels of C and N for a given set of stellar parameters, because the level of C and N can affect the thermodynamic structure of stellar atmospheres. We utilized MARCS code (Gustafsson et al. 2008) to create the specific models. Afterwards, we created synthetic spectra from the models using the Turbospectrum synthesis code (Alvarez & Plez 1998; Plez 2012). This code adopts the solar abundances from Asplund et al. (2005), the line broadening treatment by Barklem & O'Mara (1998), and the atomic lines from VALD (Kupka et al. 1999) and the literature (e.g., Hill et al. 2002; Masseron et al. 2006). The line lists of the molecular species include CH (Masseron et al. 2014), CN, and $C_2$ (Plez 2012, private communication), as well as MgH molecules from the Kurucz line lists[7].

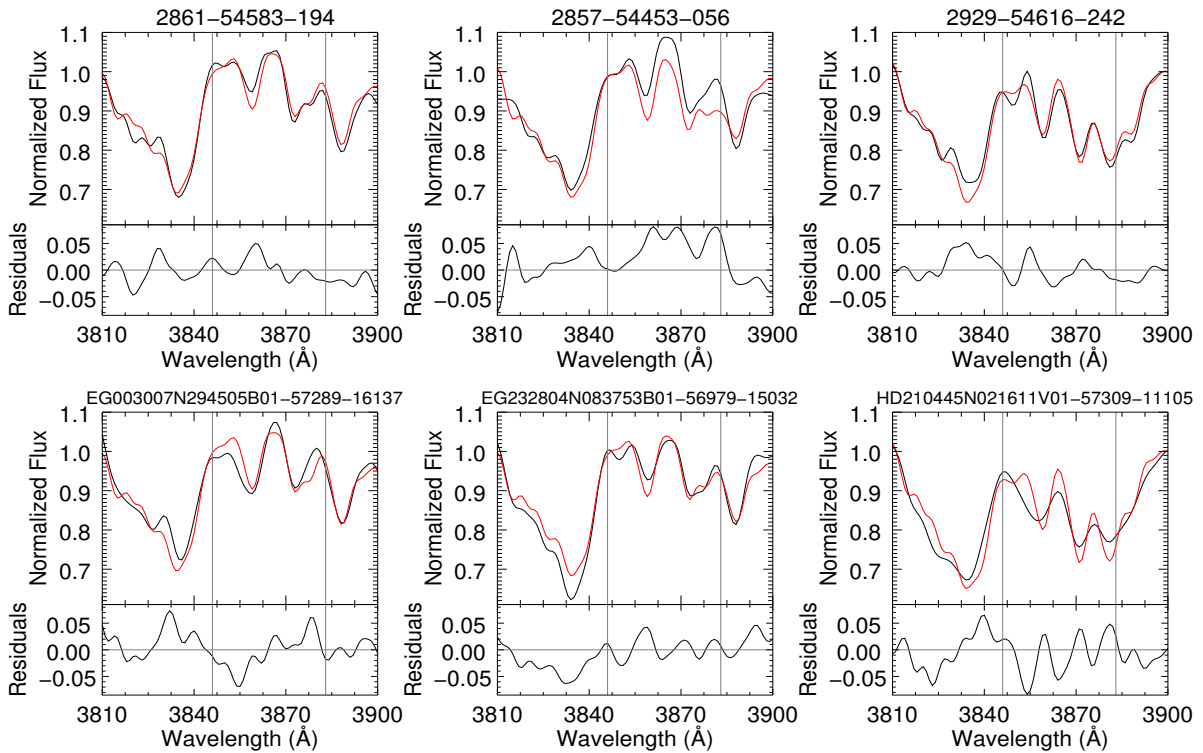

**Figure 2.** Examples of the best-matching synthetic spectra (red) produced by our approach, in the region of the molecular CN band, which is marked as vertical lines. Top panels: Three SDSS examples with similar stellar parameters and [C/Fe], but different [N/Fe], which are $T_{\rm eff}$= [5084, 5041, 5081] K, log $g$= [2.43, 2.44, 2.45], [Fe/H] = [–1.48, –1.45, –1.47], [C/Fe] = [+0.03, +0.00, +0.00], and [N/Fe] = [–0.25, +0.16, +0.72] from left to right, respectively. Bottom panels: Three LAMOST examples. Their parameters are $T_{\rm eff}$= [5073, 5063, 5048] K, log $g$= [2.49, 2.56, 2.59], [Fe/H] = [–1.54, –1.50, –1.52], [C/Fe] = [–0.02, –0.01, –0.01], and [N/Fe] = [–0.35, +0.00, +0.85] from left to right, respectively. The SDSS and LAMOST names are provided at the top of each panel. It is clear that the red synthetic spectrum well matches the black observed spectrum for all cases. Residual values between observed and synthetic spectra are plotted in the bottom panels, and are mostly less than 5%.

When creating synthetic spectra, we also considered different levels of the $\alpha$-elements (O, Mg, Si, Ca, and Ti) by following a simple linear recipe: [$\alpha$/Fe] = +0.4 for [Fe/H] < –1.0, [$\alpha$/Fe] = –0.4·[Fe/H] for –1.0 ≤ [Fe/H] ≤ 0.0, and 0.0 for [Fe/H] > 0.0, because we generally observe a similar relation among the typical disk and halo stars. Furthermore, since dwarfs and giants have different microturbulence velocities ($\xi_{\rm t}$), we adopted a simple linear relationship between microturbulence velocity and surface gravity, $\xi_{\rm t}$ [km s$^{-1}$] = 2.225 – 0.345·log $g$ to assign an appropriate $\xi_{\rm t}$ for each spectrum. This relationship was derived from the high-resolution spectra of SDSS stars used to calibrate the SSPP (Lee et al. 2008a).

The synthetic spectrum was generated at a resolving power of $R = 500,000$ over a wavelength range 3500 – 4000 Å. This spectral range is wide enough to define an accurate continuum that includes the CN band. The generated synthetic spectra cover the following ranges: 4000 K < $T_{\rm eff}$ < 7000 K in steps of 250 K, 0.0 < log $g$ < 5.0 in steps of 0.5 dex, –4.5 < [Fe/H] < +1.0 in steps of 0.25 dex, –1.5 < [C/Fe] < +4.5 in steps of 0.5 dex, and –1.5 < [N/Fe] < +4.5 in steps of 0.5 dex, resulting in a total of 555,841 spectra.

To prepare for the spectral matching, the synthetic spectra were degraded to SDSS and LAMOST spectral resolution, and linearly rebinned to 1 Å per pixel over the wavelength range 3810 – 3900 Å, in which the CN band is included. In addition, each smoothed spectrum was normalized by division of a pseudo continuum, which is obtained by the same continuum-fitting routine used for the SDSS and LAMOST spectra. Examples of degraded, normalized synthetic spectra are shown in Figure 1. In the figure, the top panels from left to right exhibit the same log $g$, [Fe/H], and [C/Fe] values of 2.5, –1.5, and 0.0, respectively, but different temperatures, 4500 K, 5000 K, and 5500 K; the [N/Fe] value changes from –0.5 (red) to +1.5 (blue) in steps of 0.5 dex. The bottom panels show the change of [C/Fe]. We clearly see in the figure that the CN-band strength increases with increasing [N/Fe] ([C/Fe]) values at fixed stellar parameters and [C/Fe] ([N/Fe]). We also note that the CN band is very sensitive to the temperature, as its strength becomes very weak at $T_{\rm eff}$ = 5500 K and above.

---

[7] http://kurucz.harvard.edu

### 2.3.2. Measurement of [N/Fe]

We need to pre-process the SDSS and LAMOST spectra before attempting to determine [N/Fe]. The pre-processing steps are as follows. First, as the wavelength scale is in vacuum, we changed it to an air-based scale and adjusted the spectrum to the rest frame with a radial velocity determined from the spectrum. We then re-sampled flux values in the wavelength range 3810 – 3900 Å, and normalized the observed spectrum by the following procedure. The first step was to calculate the median values of the fluxes over 3810 – 3814 Å and 3896 – 3900 Å wavelength ranges, respectively, in which strong absorption features do not exist. The next step was to obtain a pseudo continuum by connecting the two median values. The normalized spectrum was obtained by dividing the spectrum by this continuum.

Once the normalized spectrum was ready, we searched the grid of synthetic spectra for the best-fitting model parameters by minimizing the difference between the normalized observed spectrum, $O$, and the normalized synthetic spectrum, $S$, over the wavelength range 3810 – 3900 Å, using a $\chi^2$ approach, given by:

$$\chi^2 = \sum_{i=1}^{n} \frac{(O_i - S_i)^2}{\sigma_i^2} \quad (1)$$

where $\sigma_i$ is the error in flux in the $i$th pixel. Each SDSS or LAMOST spectrum contains the error in the flux associated with each pixel. The IDL MPFIT routine (Markwardt 2009) was used to find the minimum $\chi^2$ value at which the [N/Fe] value is determined. While determining [N/Fe], we adopted $T_{\rm eff}$, log $g$, [Fe/H], and [C/Fe] values from the SSPP and held these values fixed, and varied only the [N/Fe] value to produce a new trial synthetic spectrum by cubic spline interpolation over the grid of the synthetic spectra. During the tests of this technique, we realized that fixing $T_{\rm eff}$, log $g$, [Fe/H], and [C/Fe] constant allowed for more subtle variations of [N/Fe] to be explored. The fitting error comes from the co-variance (which is the standard deviation) of the altered [N/Fe] values during the minimization.

**Table 1**
Impact of S/N on Estimated [N/Fe]

| S/N | $\Delta$[N/Fe] (Ours - Model) [dex] | | | | | |
| --- | --- | --- | --- | --- | --- | --- |
| | Giant | | Dwarf | | Turnoff | |
| | $\mu$ | $\sigma$ | $\mu$ | $\sigma$ | $\mu$ | $\sigma$ |
| 10 | 0.017 | 0.034 | 0.035 | 0.048 | 0.018 | 0.038 |
| 20 | 0.004 | 0.016 | 0.008 | 0.022 | 0.001 | 0.021 |
| 30 | 0.001 | 0.011 | 0.003 | 0.015 | –0.001 | 0.013 |
| 40 | 0.001 | 0.009 | 0.001 | 0.012 | –0.001 | 0.010 |
| 50 | 0.001 | 0.008 | 0.001 | 0.010 | 0.001 | 0.007 |

$\mu$ and $\sigma$ denote the Gaussian mean and standard deviation. See the text for the selection criteria for each group of stars.

Figure 2 shows six examples (three from SDSS and three from LAMOST) of best-matching synthetic spectra determined by our approach. The stars shown in Figure 2 have similar $T_{\rm eff}$, log $g$, [Fe/H], and [C/Fe] as listed in the caption of the figure. The [N/Fe] values are –0.25, +0.16, and +0.72 in the upper panels and –0.35, 0.00, +0.85 in the bottom panels, from left to right. In this figure, we see good matches between the synthetic and observed spectrum, proving that our approach performs well enough to identify the N-enhanced stars even for low-resolution spectra. The distribution of residuals shown in the bottom of each panel suggests that the largest deviation is no more than 10%; they are mostly less than 5%.

## 3. CHECKS ON THE ROBUSTNESS OF ESTIMATED [N/FE]
### 3.1. Impacts of S/N

The S/N ratios of stellar spectra obtained by SDSS and LAMOST have a very wide range. Because the noise in a spectrum can play a role in producing false signals, it could cause large systematic deviations and random scatter for the chemical abundances to be determined. For this reason, it is very important to examine the effects of S/N on the estimated abundances of [N/Fe]. In order to appreciate the impacts of the S/N on our derived [N/Fe], we injected different levels of random noise, which simulates the flux error in an SDSS spectrum into the synthetic spectra used to estimate [N/Fe], following Lee et al. (2008a). We constructed 10 different noise-injected synthetic spectra for a given model spectrum at a given S/N ratio of 10, 20, 30, 40, and 50, and then applied our technique to these noise-injected spectra to reproduce the model value of [N/Fe]. While minimizing the $\chi^2$ values, we fixed $T_{\rm eff}$, log $g$, [Fe/H], and [C/Fe] values associated with each synthetic spectrum held constant, and altered only the [N/Fe] value to minimize the $\chi^2$ values. Note that, even if we increase the number of the noise-injected synthetic spectra from 10 to 100, the following results of our exercise do not change much.

Figure 3 shows the results of our noise-injection exercise. In this figure, we examine how the S/N affects the determined [N/Fe] in the plane of $T_{\rm eff}$ versus S/N, log $g$ versus S/N, [Fe/H] versus S/N, and [C/Fe] versus S/N for three different groups of stars: giants, main-sequence turnoff stars, and dwarfs, from top to bottom.

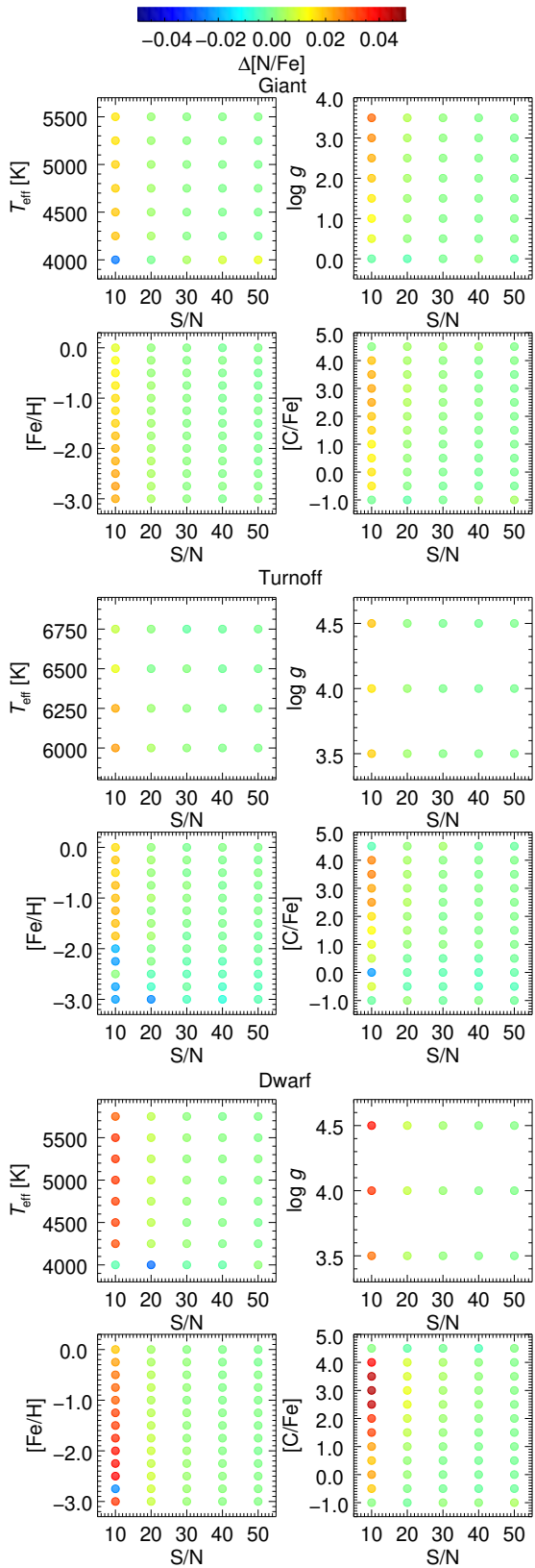

**Figure 3.** Color-coded residual distribution of [N/Fe] between ours and model values for three groups of stars (giants, main-sequence turnoff stars, and dwarfs). See the text for the selection criteria for each group of stars. The scale of the differences is displayed as a color bar at the top. Each panel shows the [N/Fe] difference in the plane of stellar parameters or [C/Fe] versus S/N.

The giant group corresponds to the ranges $T_{\rm eff} \leq 5500$ K and $\log g \leq 3.5$, the turnoff group spans the ranges $T_{\rm eff} \geq 6000$ K and $\log g \geq 3.5$, and the dwarf group covers the ranges $T_{\rm eff} < 6000$ K and $\log g \geq 3.5$. We refer to the parameter space of the isochrones by Demarque et al. (2004) to divide into the three groups. When dividing into these three groups, we excluded the model spectrum with [Fe/H] < −3.0, because such objects are very rare, and we are mainly interested in the discovery of GC-origin stars or the debris from already-disrupted GCs, which are likely to have [Fe/H] > −3.0[8]. The considered range of [N/Fe] is −1.0 to +4.5.

The color-coded circles denote the mean difference in [N/Fe] between our estimates and the model values, which were computed by considering all noise-injected spectra with the given S/N, but not with the given parameter in the ordinate in each panel. The scale of the differences is displayed as a color bar at the top. Although Figure 3 indicates no large mean differences, we do observe that the differences increase with decreasing S/N for all parameters in the three groups, as expected. This pattern is more manifest for the dwarf population. The maximum difference, however, still does not exceed $\sim 0.05$ dex.

A more quantitative analysis is provided in Table 1. In this table, we list the Gaussian mean offsets and scatter, which are calculated from all [N/Fe] estimates within the indicated S/N. We clearly see that the mean offsets and standard deviations tend to increase as the S/N decreases. However, their size is relatively small, mostly less than 0.05 dex. We conclude from the behaviors seen in Figure 3 and Table 1 that the S/N does not greatly worsen the accuracy and precision of our estimated [N/Fe]. Based on this, even though we can set the S/N limit for our technique to S/N ≥ 10, we conservatively recommend the application of this technique to spectra with S/N ≥ 15 per Å.

### 3.2. Impacts of Systematic Errors of Adopted Stellar Parameters on Determined [N/Fe]

While searching for a best-matching synthetic spectrum via the $\chi^2$ minimization, we do not alter the $T_{\rm eff}$, $\log g$, [Fe/H], and [C/Fe] values delivered by the SSPP, but only [N/Fe]. Given that the strength of the CN band used for measuring [N/Fe] is sensitive to the change in these parameters, any systematic errors in those parameters leads to inaccurate measurement of [N/Fe]. Thus, we need to understand how the systematic errors in the input parameters affect the accuracy and precision of the measured [N/Fe]. We assessed these effects using the grid of synthetic spectra once again. To begin with, we fed a synthetic spectrum as input into our routine to measure [N/Fe]. While estimating [N/Fe], we shifted its model values by ±100 K for $T_{\rm eff}$ and ±0.2 dex for $\log g$, [Fe/H], and [C/Fe], which are assumed to be typical uncertainties of the SSPP for a spectrum with a decent S/N (∼ 30).

---

[8] Note that a recent study (Martin et al. 2022) reported possible candidates of GC-origin stars with [Fe/H] < −3.0.

**Table 2**
Effects of Stellar Parameter and [C/Fe] Errors on Estimated [N/Fe]

| Type | $T_{\rm eff}$ Shift (K) | $\Delta$[N/Fe] $\mu$ (dex) | $\sigma$ (dex) | log $g$ Shift (dex) | $\Delta$[N/Fe] $\mu$ (dex) | $\sigma$ (dex) | [Fe/H] Shift (dex) | $\Delta$[N/Fe] $\mu$ (dex) | $\sigma$ (dex) | [C/Fe] Shift (dex) | $\Delta$[N/Fe] $\mu$ (dex) | $\sigma$ (dex) |
|---|---|---|---|---|---|---|---|---|---|---|---|---|
| Giant   | +100 | +0.144 | 0.255 | +0.2 | +0.082 | 0.226 | +0.2 | −0.260 | 0.288 | +0.2 | −0.120 | 0.226 |
| Dwarf   | +100 | +0.079 | 0.238 | +0.2 | +0.110 | 0.220 | +0.2 | −0.216 | 0.301 | +0.2 | −0.083 | 0.262 |
| Turnoff | +100 | +0.012 | 0.516 | +0.2 | −0.127 | 0.469 | +0.2 | −0.303 | 0.428 | +0.2 | −0.188 | 0.434 |
| Giant   | −100 | −0.134 | 0.253 | −0.2 | −0.086 | 0.250 | −0.2 | +0.259 | 0.315 | −0.2 | +0.123 | 0.297 |
| Dwarf   | −100 | −0.086 | 0.248 | −0.2 | −0.120 | 0.201 | −0.2 | +0.203 | 0.326 | −0.2 | +0.098 | 0.321 |
| Turnoff | −100 | −0.217 | 0.452 | −0.2 | −0.084 | 0.459 | −0.2 | +0.102 | 0.535 | −0.2 | −0.018 | 0.459 |

The shift in each parameter means the deviation from the model value of a synthetic spectrum while determining [N/Fe]. $\mu$ and $\sigma$ are the mean difference and standard deviation derived by a Gaussian fit, between the estimated [N/Fe] with the perturbed parameter and the original model value. See the text for selecting each group of stars.

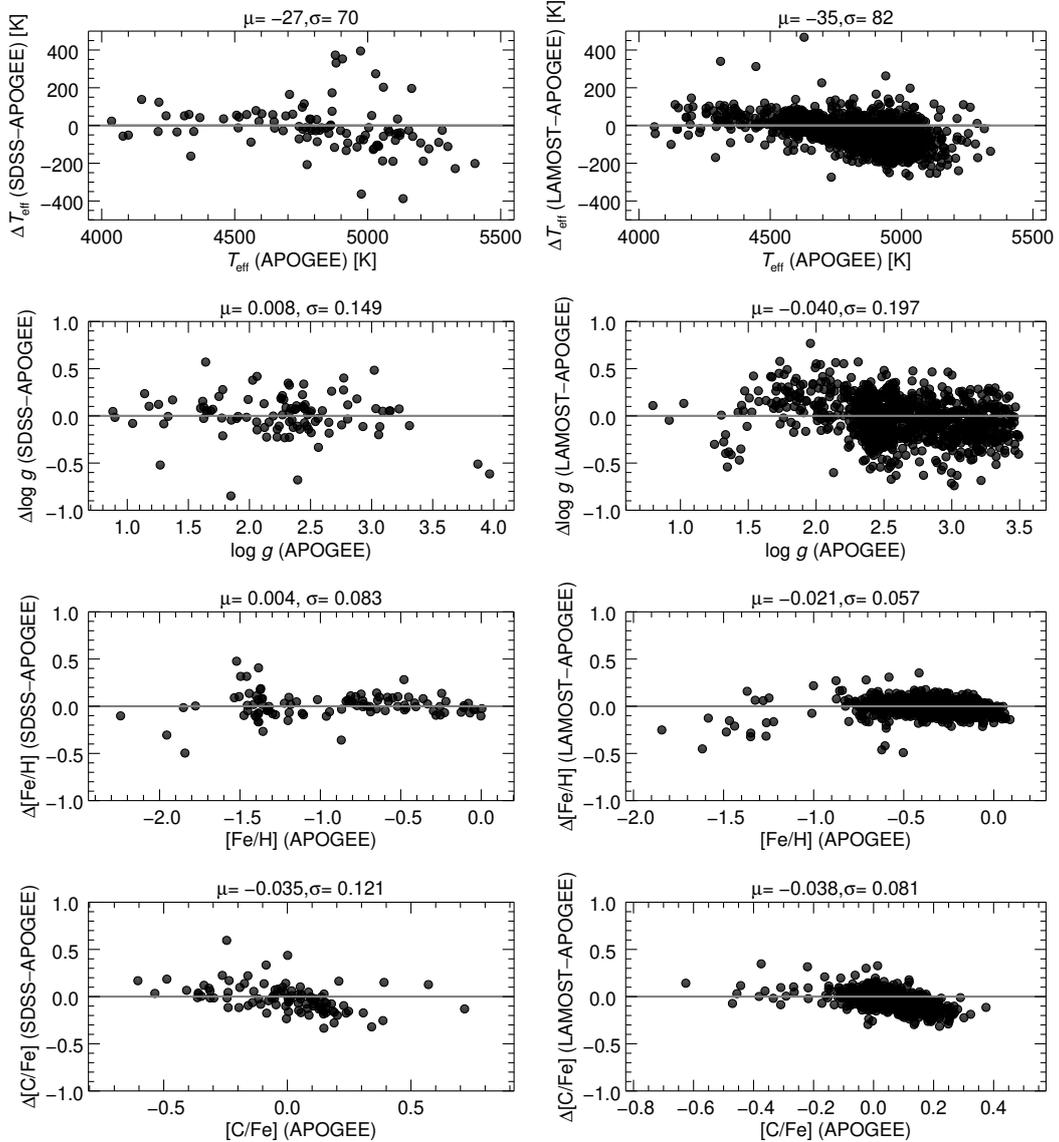

**Figure 4.** Differences in stellar parameters and [C/Fe] between SDSS and APOGEE (left panels), and LAMOST and APOGEE (right panels), as a function of $T_{\rm eff}$, log $g$, [Fe/H], and [C/Fe], from top to bottom. The mean offsets and standard deviations are indicated at the top of each panel. Even though we do not find any substantial systematic behaviors for $T_{\rm eff}$, log $g$, and [Fe/H] in either survey data, we observe a minor trend for [C/Fe] > +0.1. See the text for the possible reason for this.

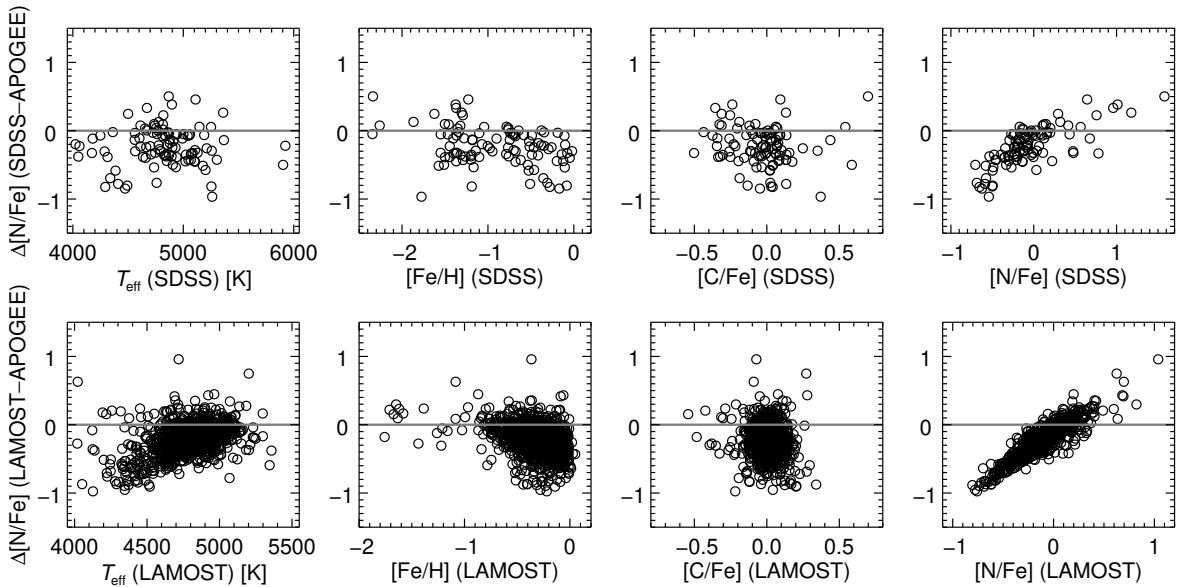

**Figure 5.** Comparison of our estimated [N/Fe] from SDSS (top panels) and LAMOST (bottom panels) with that of APOGEE, as a function of $T_{\rm eff}$, [Fe/H], [C/Fe], or [N/Fe]. The systematic trends for both SDSS and LAMOST are clearly present. We applied a linear regression as functions of $T_{\rm eff}$, [Fe/H], and [N/Fe] to derive a correction function to remove the systematic trends, as described in the text.

After estimating [N/Fe], we divided the determined [N/Fe] values into giant, turnoff, and dwarf groups as in Section 3.1, and investigated the collective patterns of [N/Fe] for each group. The results of this exercise are summarized in Table 2. This table lists the Gaussian mean offsets and scatters for each group for each parameter shift. From inspection, we note that all but [Fe/H] show a mean offset mostly less than 0.2 dex for the three samples. For the case of [Fe/H], the mean offset goes up to 0.3 dex. For giants and dwarfs, most of the scatters are less than 0.3 dex for all parameters. For the case of the turnoff group, the standard deviation increases up to 0.5 dex. This is primarily due to their high effective temperatures, which weaken the CN band. This exercise indicates that the metallicity is the most sensitive parameter, which can cause the measured [N/Fe] to be systematically different to a larger extent than the other parameters. For the turnoff stars with high temperatures, the systematic error in the SSPP $T_{\rm eff}$ is sensitive as well.

Among the three groups, we are more interested in the estimation of [N/Fe] for the giant group. This is because we expect to find only bright halo giants as candidates of the GC-origin stars in the SDSS and LAMOST stellar databases. The main-sequence turnoff stars can be observed, but due to their high temperatures, it is rather difficult to measure a reliable [N/Fe], as shown in Figure 3 and Table 2. We conclude that, given that the error of the abundance of a chemical element (such as [Fe/H] or [C/Fe]) that can be measured from low-resolution spectra is generally about 0.3 dex, the additional error in [N/Fe] caused by the systematic errors of the SSPP parameters is not significant. This is in particular the case for the giants, which we are most interested in.

## 4. CALIBRATION AND VALIDATION WITH A HIGH-RESOLUTION ABUNDANCE ANALYSIS

Having developed a technique for the estimation of [N/Fe], and carried out some tests to check on the robustness of its measurement, we turn to procedures for calibrating and validating our method, based on a star-by-star comparison with stars having external measurements. In this manner, we seek to quantify any systematic offsets and effectively eliminate them, as well as estimate the random scatter associated with the estimated [N/Fe]. For these calibration and validation exercises, we introduce the [N/Fe] measurements based on high-resolution spectroscopy from the APOGEE and SAGA databases, respectively.

### 4.1. Checks on the Accuracy and Precision of the SSPP Stellar Parameters and [C/Fe]

Because we adopt the SSPP stellar parameters and [C/Fe] when we determine [N/Fe], we need to check on the reliability of the SSPP stellar parameters first before we move to the calibration of our derived [N/Fe]. The APOGEE survey obtained near-infrared spectra in the H band at a resolving power of $R \sim 25{,}000$ and provides accurate stellar parameters and [C/Fe] for a large number of stars, hence it is useful for this comparison exercise. Note that, because we are more interested in identifying N-rich halo giants for the follow-up study, this comparison and our calibration effort in the following section are restricted to the giants.

We cross-matched SDSS DR12 (Alam et al. 2015) and LAMOST DR5 (Luo et al. 2019) stellar sources with APOGEE data in the SDSS DR16 (DR16; Ahumada et al. 2020) within $1''$, and selected all RGB stars with [N/Fe] available in the ranges 4000 K < $T_{\mathrm{eff}}$ < 6000 K, 0 < log $g$ < 3.5, and −2.5 < [Fe/H] < 0.0. We also made sure that the "ASPCAPFLAG" flag is set to 0, to remove any problematic stars with the flags such as poor synthetic spectral fit and stellar parameters estimated near grid boundaries. In addition, we eliminated the stars with errors in the determined [N/Fe] larger than 0.2 dex in the APOGEE data, or with radial-velocity differences between the two surveys larger than 10 km s$^{-1}$. Through this selection procedure, we identified 100 acceptable stars in common between SDSS and APOGEE, and 1058 acceptable stars between LAMOST and APOGEE, respectively.

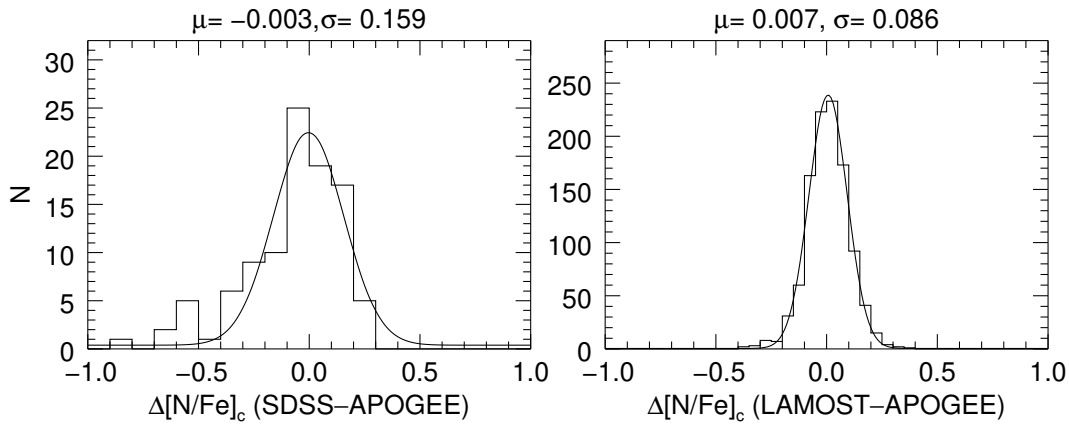

**Figure 6.** Residual distributions in $[\mathrm{N/Fe}]_c$, after applying the derived correction functions. We see very small offsets and a standard deviation of 0.159 dex and 0.086 dex for SDSS and LAMOST, respectively, which are derived from a Gaussian fit to the residuals.

Using these selected samples, we have assessed the accuracy and precision of the SSPP stellar parameters and [C/Fe]. Figure 4 illustrates the results of the comparisons with the APOGEE values. Except for [C/Fe], which exhibits a slight minor trend for [C/Fe] > +0.1 or so, we do not notice any significant trends for the other parameters. The reason for this minor trend may stem from the use of the different molecular bands to determine [C/Fe]. In APOGEE, the carbon abundance ratio is determined from the CO band after the oxygen abundance ratio is estimated from the OH band. Because the CO molecule has very strong bond, if the C abundance is larger than the O abundance, most of the oxygen is locked up in CO and other oxygen compounds such as OH are absent or too weak to determine the O abundance ratio, resulting in the large uncertainty in the C abundance ratio. On the other hand, our [C/Fe] is determined from the CH band around 4300 Å. Consequently, we infer that the use of the different molecular bands for the carbon indicator causes the mild systematic trend. Nonetheless, the systematic offset is almost zero and the scatter is very small, mostly less than 0.2 dex for log $g$, [Fe/H], and [C/Fe], and 100 K for $T_{\mathrm{eff}}$. They are much lower than the typical uncertainties of 0.3 dex and 200 K from the low-resolution spectroscopic analysis. This comparison ensures that the SSPP estimates stellar parameters and [C/Fe] with good accuracy and precision from the low-resolution SDSS and LAMOST spectra.

### 4.2. Calibration with APOGEE

Although the [N/Fe] values reported in the APOGEE DR16 are determined from different CN features in the wavelength range 15,260 – 15,482 Å (Smith et al. 2013), because we expect that the estimated [N/Fe] should be the same regardless of the lines analyzed, it is worthwhile to carry out this calibration effort with the APOGEE data. Another reason for adopting the APOGEE data to compare with is that the [N/Fe] value determined from the APOGEE data is based on much higher spectral resolution, which is less affected by other lines, resulting in more accurate [N/Fe].

The top panels of Figure 5 display the initial comparison results for the SDSS results; the bottom panels for the LAMOST results. We note some systematic trends of the [N/Fe] difference as a function of $T_{\mathrm{eff}}$, [Fe/H], [C/Fe], and [N/Fe]. It appears that the systematic trend is similar for both survey data. We suspect that the origin of these systematic offsets involves poor flux calibration of the low-resolution data in the blue spectral region. In spectroscopic surveys, F-type stars are often used to carry out the flux calibration. The figure implies that most of cool stars contribute to the systematic trend. These stars have relatively lower S/N than the F-type calibration stars in the CN band, resulting in the poor flux calibration for the cool stars. In turn, the strength of the CN band is not well-reflected in the poorly flux calibrated spectrum, causing the systematic offset observed in the figure. Regardless of its origin, to remove these systematic trends, we derived a correction function by a regression analysis as functions of $T_{\mathrm{eff}}$, [Fe/H], and [N/Fe] for each dataset, given by:

1. SDSS

$$[\mathrm{N/Fe}]_c = [\mathrm{N/Fe}] - (-0.191 - 0.09 \cdot 10^{-4} \cdot T_{\mathrm{eff}} \\ - 0.036 \cdot [\mathrm{Fe/H}] + 0.505 \cdot [\mathrm{N/Fe}]) \quad (2)$$

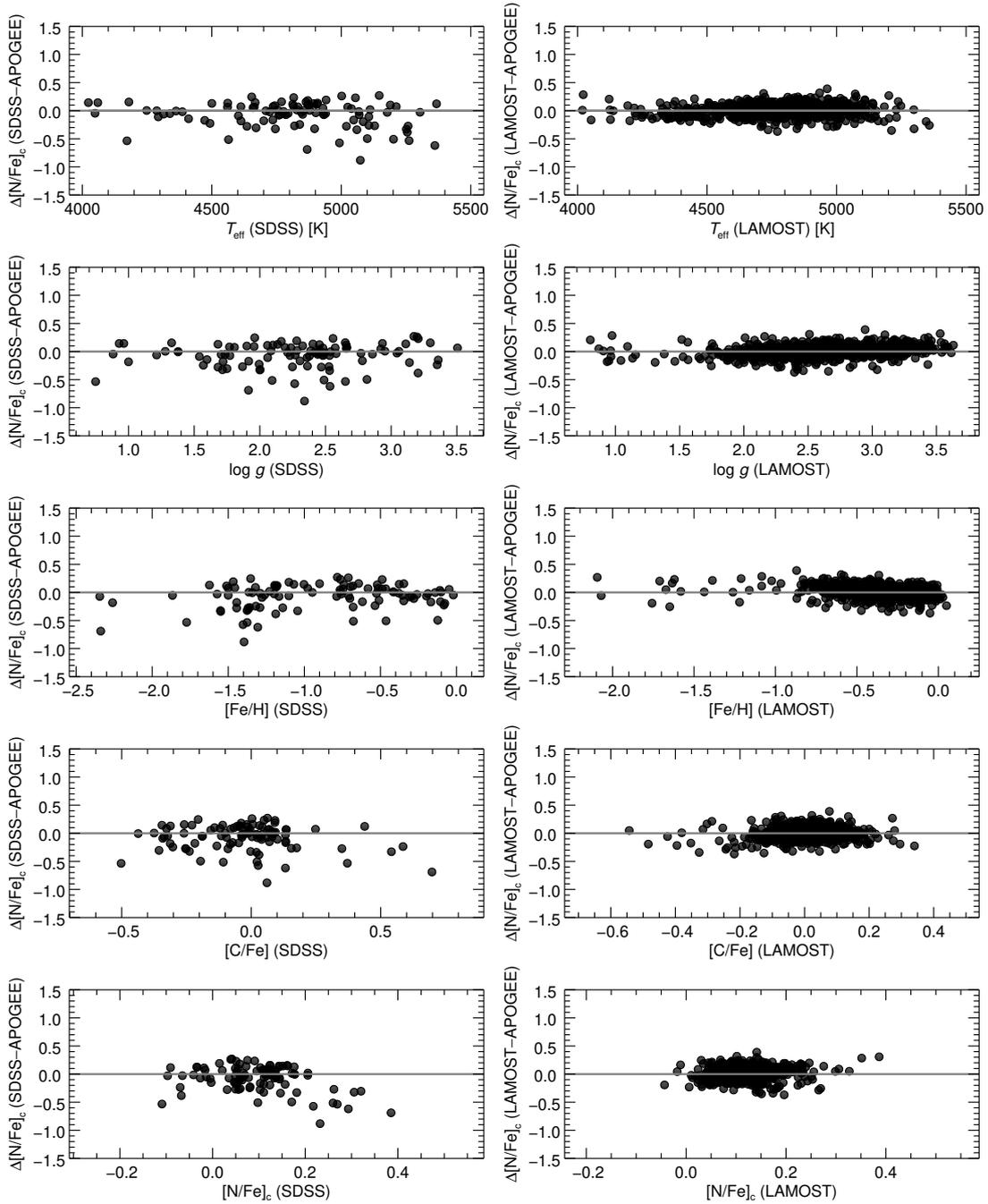

**Figure 7.** Difference in $[N/Fe]_c$ between our estimates and APOGEE values, as a function of $T_{\rm eff}$, log $g$, [Fe/H], [C/Fe], and $[N/Fe]_c$, after applying the correction function. The left panels are for SDSS, and the right panels for LAMOST. We do not notice any significant systematic trends.

2. LAMOST

$$[N/Fe]_c = [N/Fe] - (-0.859 + 1.41 \cdot 10^{-4} \cdot T_{\rm eff} \\ - 0.155 \cdot [Fe/H] + 0.840 \cdot [N/Fe]) \quad (3)$$

where $T_{\rm eff}$, [Fe/H], and [N/Fe] are the SSPP values derived from the SDSS or LAMOST spectra and $[N/Fe]_c$ is the corrected estimate. We note in Figure 5 that the systematic trend over [C/Fe] is smaller than for other other parameters; hence there is no [C/Fe] term in the above equations. In addition, we derived separate correction function for SDSS and LAMOST, because we obtained a better behavior in the estimated [N/Fe] with a separated correction function for each survey than a correction function derived from the combined data of SDSS and LAMOST.

Figure 6 exhibits the residual distributions in $[N/Fe]_c$ after removing the systematic trends using the Equations 2 and 3, over-plotted with a fitted Gaussian. We see almost zero offset and a standard deviation of 0.159 dex and 0.086 dex for SDSS and LAMOST, respectively, derived by a Gaussian fit to the residuals.

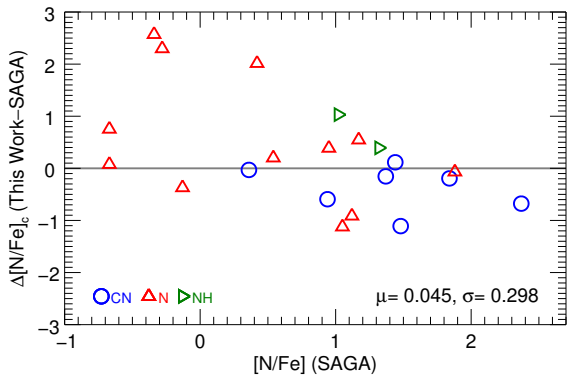

**Figure 8.** Comparison of our estimated [N/Fe]$_c$ with that of a small number of stars with high-resolution estimates from the SAGA database. Blue circles indicate [N/Fe] estimated from the CN features, red triangles from the atomic N lines, and the green triangles from the NH band. The gray solid line is the zero point, and $\mu$ and $\sigma$ are the Gaussian mean and standard deviation of the differences.

Once again, we examined any systematic trend of our derived [N/Fe]$_c$, as a function of $T_{\rm eff}$, log $g$, [Fe/H], [C/Fe], and [N/Fe], in Figure 7 for both SDSS (left panels) and LAMOST (right panels), and found negligible trends with the five parameters for both survey samples, suggesting that our calibration works well. Therefore, we decided to apply the derived correction functions to obtain the estimated [N/Fe]$_c$ and to put it on the same abundance scale as the APOGEE survey data.

### 4.3. Validation with SAGA

The SAGA database provides chemical information for stars that were analyzed with high-resolution spectroscopy, compiled from various studies. By cross-matching the SAGA database with our SDSS and LAMOST stars, we found 25 SDSS and LAMOST stars with [N/Fe] available in the parameter ranges that we are interested in, as described in Section 4.1. The [N/Fe] value in the SAGA catalog is derived from the CN and NH molecular features or atomic N lines. We further checked the quality of the spectra of the selected stars, and found that four stars were observed in LAMOST and have poor flux calibration around the CN band. The quality of the flux calibration in the LAMOST spectra is relatively poorer than for the SDSS spectra in general. We decided to exclude these stars from the comparison sample, which leaves a total of 21 objects to compare with the SAGA.

Figure 8 shows the differences in [N/Fe] between ours and the SAGA values. In the figure, our estimate of [N/Fe] is already adjusted with the correction functions derived in Section 4.2. The blue circles indicate [N/Fe] estimated from the CN features, while the red triangles from the atomic N lines. The green triangles denote the [N/Fe] determination from the NH band. Note that we observe three stars with very large mean offsets (> 2.0 dex). We found that their temperature (or gravity) estimates differ by more than 500 K (or 0.75 dex) from ours, which causes the large systematic offsets from our estimate of [N/Fe]. We could not identify the reason for the large difference, however. We already noted in Figure 1 that the nitrogen abundance ratio is very sensitive to temperature, and Collet et al. (2007) also demonstrated with their 3D hydrodynamical models that the molecular line strength is sensitive to $T_{\rm eff}$ for cool stars.

Even with the inclusion of the three objects, our derived mean and standard deviation from the Gaussian fit indicate 0.045 dex and 0.298 dex, respectively, which are not far from those we found from the APOGEE comparison. The comparisons with both APOGEE and SAGA data assure that our [N/Fe] estimate is in good agreement with that from high-resolution spectroscopic analyses. As a result, we conclude that our method can be applied to low-resolution stellar spectra to identify the N-rich (especially giant) stars in the Galactic halo.

### 5. SUMMARY AND CONCLUSIONS

We have presented a method for measuring [N/Fe] from SDSS and LAMOST low-resolution spectra via spectral matching with a grid of synthetic spectra around the molecular CN band. We carried out several tests to investigate the robustness of our derived [N/Fe], and summarize the results of these tests below.

The noise-injection test indicates that, as the S/N decreases, the systematic and random errors increase, as expected. However, their size are less than 0.05 dex, proving we can obtain robust estimates of [N/Fe] even with poor-S/N spectra. The test for the impact of the systematic errors in the adopted stellar parameters ($T_{\rm eff}$, log $g$, [Fe/H], and [C/Fe]) suggests that both the systematic and random errors are less than 0.3 dex for giant and dwarf stars, but not for turnoff stars, which exhibit a larger scatter, up to 0.5 dex. Nonetheless, since we are more interested in the halo giants for follow-up research, this test confirms that we can use our approach to identify N-enhanced stars from low-resolution SDSS and LAMOST stellar spectra.

Our estimated [N/Fe] was compared and calibrated with the APOGEE values to remove systematic trends with $T_{\rm eff}$, [Fe/H], and [N/Fe]. After calibration, we obtained negligible mean offsets and scatters less than 0.2 dex for both SDSS and LAMOST. This level of precision is confirmed by the comparison with about a dozen of stars in the SAGA database.

Through these various tests, we conclude that our method to estimate [N/Fe] has sufficient accuracy and precision to identify N-rich stars among the giants in the Galactic halo, using low-resolution SDSS and LAMOST stellar spectra. In a follow-up study, we will attempt to identify the N-enhanced stars from the large spectroscopic survey data, and study their association with dissolved GC stars.

### ACKNOWLEDGMENTS

We thank anonymous referees for a careful review of this paper, which improved the clarity of its presen-tation. This work was supported by research fund of Chungnam National University. T. C. B. acknowledges partial support for this work from grant PHY 14-30152; Physics Frontier Center/JINA Center for the Evolution of the Elements (JINA-CEE), awarded by the U.S. National Science Foundation. T. M. acknowledges financial support from the Spanish Ministry of Science and Innovation (MICINN) through the Spanish State Research Agency, under the Severo Ochoa Program 2020-2023 (CEX2019-000920-S).

Funding for the Sloan Digital Sky Survey IV has been provided by the Alfred P. Sloan Foundation, the U.S. Department of Energy Office of Science, and the Participating Institutions.

SDSS-IV acknowledges support and resources from the Center for High Performance Computing at the University of Utah. The SDSS website is www.sdss.org.

SDSS-IV is managed by the Astrophysical Research Consortium for the Participating Institutions of the SDSS Collaboration including the Brazilian Participation Group, the Carnegie Institution for Science, Carnegie Mellon University, Center for Astrophysics — Harvard & Smithsonian, the Chilean Participation Group, the French Participation Group, Instituto de Astrofísica de Canarias, The Johns Hopkins University, Kavli Institute for the Physics and Mathematics of the Universe (IPMU) / University of Tokyo, the Korean Participation Group, Lawrence Berkeley National Laboratory, Leibniz Institut für Astrophysik Potsdam (AIP), Max-Planck-Institut für Astronomie (MPIA Heidelberg), Max-Planck-Institut für Astrophysik (MPA Garching), Max-Planck-Institut für Extraterrestrische Physik (MPE), National Astronomical Observatories of China, New Mexico State University, New York University, University of Notre Dame, Observatário Nacional / MCTI, The Ohio State University, Pennsylvania State University, Shanghai Astronomical Observatory, United Kingdom Participation Group, Universidad Nacional Autónoma de México, University of Arizona, University of Colorado Boulder, University of Oxford, University of Portsmouth, University of Utah, University of Virginia, University of Washington, University of Wisconsin, Vanderbilt University, and Yale University.

The Guoshoujing Telescope (the Large Sky Area Multi- Object Fiber Spectroscopic Telescope, LAMOST) is a National Major Scientific Project which is built by the Chinese Academy of Sciences, funded by the National Development and Reform Commission, and operated and managed by the National Astronomical Observatories, Chinese Academy of Sciences.

## References

Abolfathi, B., Aguado, D. S., Aguilar, G., et al. 2018, The Fourteenth Data Release of the Sloan Digital Sky Survey: First Spectroscopic Data from the Extended Baryon Oscillation Spectroscopic Survey and from the Second Phase of the Apache Point Observatory Galactic Evolution Experiment, ApJS, 235, 42

Ahumada, R., Prieto, C. A., Almeida, A., et al. 2020, The 16th Data Release of the Sloan Digital Sky Surveys: First Release from the APOGEE-2 Southern Survey and Full Release of eBOSS Spectra, ApJS, 249, 3

Alam, S., Albareti, F. D., Allende Prieto, C., et al. 2015, The Eleventh and Twelfth Data Releases of the Sloan Digital Sky Survey: Final Data from SDSS-III, ApJS, 219, 12

Allende Prieto, C., Beers, T. C., Wilhelm, R., et al. 2006, A Spectroscopic Study of the Ancient Milky Way: F- and G-Type Stars in the Third Data Release of the Sloan Digital Sky Survey, ApJ, 636, 804

Allende Prieto, C., Sivarani, T., Beers, T. C., et al. 2008, The SEGUE Stellar Parameter Pipeline. III. Comparison with High-Resolution Spectroscopy of SDSS/SEGUE Field Stars, AJ, 136, 2070

Alvarez, R., & Plez, B. 1998, Near-infrared narrow-band photometry of M-giant and Mira stars: models meet observations, A&A, 330, 1109

Asplund, M., Grevesse, N., & Sauval, A. J. 2005, The Solar Chemical Composition, Cosmic Abundances as Records of Stellar Evolution and Nucleosynthesis, 336, 25

Bailer-Jones, C. A. L. 2000, Stellar parameters from very low resolution spectra and medium band filters. $T_{eff}$, log g and [M/H] using neural networks, A&A, 357, 197

Barklem, P. S., & O'Mara, B. J. 1998, The broadening of strong lines of Ca+, Mg+ and Ba+ by collisions with neutral hydrogen atoms, MNRAS, 300, 863

Bastian, N., & Lardo, C. 2018, Multiple Stellar Populations in Globular Clusters, ARA&A, 56, 83

Belokurov, V., Erkal, D., Evans, N. W., Koposov, S. E., & Deason, A. J. 2018, Co-formation of the disc and the stellar halo, MNRAS, 478, 611

Carollo, D., Martell, S. L., Beers, T. C., & Freeman, K. C. 2013, CN Anomalies in the Halo System and the Origin of Globular Clusters in the Milky Way, ApJ, 769, 87

Carretta, E., Bragaglia, A., Gratton, R. G., et al. 2009, Na-O anticorrelation and HB. VII. The chemical composition of first and second-generation stars in 15 globular clusters from GIRAFFE spectra, A&A, 505, 117

Carretta, E. 2016, Globular clusters and their contribution to the formation of the Galactic halo, The General Assembly of Galaxy Halos: Structure, Origin and Evolution, 317, 97

Collet, R., Asplund, M., & Trampedach, R. 2007, Three-dimensional hydrodynamical simulations of surface convection in red giant stars, A&A, 469, 687

Cui, X.-Q., Zhao, Y.-H., Chu, Y.-Q., et al. 2012, The Large Sky Area Multi-Object Fiber Spectroscopic Telescope (LAMOST), RAA, 12, 1197

Demarque, P., Woo, J.-H., Kim, Y.-C., & Yi, S. K. 2004, $Y^2$ Isochrones with an Improved Core Overshoot Treatment, ApJS, 155, 667

Fernandez-Trincado, J. G., Zamora, O., Souto, D., et al. 2019a, H-Band Discovery of Additional Second-Generation Stars in the Galactic Bulge Globular Cluster NGC 6522 as Observed by APOGEE and Gaia, A&A, 627, 178

Fernandez-Trincado, J. G., Beers, T. C., Placco, V. M., et al. 2019b, Discovery of a New Stellar Sub-Population Residing in the (Inner) Stellar Halo of the Milky Way, ApJ, 886, L8

Fernandez-Trincado, J. G., Minniti, D., Beers, T. C., , et al. 2020a, The Enigmatic Globular Cluster UKS 1 Obscured by the Bulge: H-band Discovery of Nitrogen-Enhanced Stars, A&A, 643, 145

Fernandez-Trincado, J. G., Beers, T. C., Minniti, D., et al. 2020b, Discovery of a Large Population of Nitrogen-Enhanced Stars in the Magellanic Clouds, ApJ, 903, L17

Fernandez-Trincado, J. G., Beers, T. C., Minniti, D., et al. 2021a, APOGEE Discovery of a Chemically Atypical Star Disrupted from NGC 6723 and Captured by the Milky Way Bulge, A&A, 647, 64

Fernandez-Trincado, J. G., Beers, T. C., Queiroz, A. B. A., et al. 2021b, APOGEE-2 Discovery of a Large Population of Relatively High-Metallicity Globular Cluster Debris, ApJ, 918, L37

Gaia Collaboration, Brown, A. G. A., Vallenari, A., et al. 2018, Gaia Data Release 2. Summary of the contents and survey properties, A&A, 616, A1

Gustafsson, B., Edvardsson, B., Eriksson, K., et al. 2008, A grid of MARCS model atmospheres for late-type stars. I. Methods and general properties, A&A, 486, 951

Helmi, A., Babusiaux, C., Koppelman, H. H., et al. 2018, The merger that led to the formation of the Milky Way's inner stellar halo and thick disk, NATURE, 563, 85

Hill, V., Plez, B., Cayrel, R., et al. 2002, First stars. I. The extreme r-element rich, iron-poor halo giant CS 31082-001. Implications for the r-process site(s) and radioactive cosmochronology, A&A, 387, 560

Horta, D., Mackereth, J. T., Schiavon, R. P., et al. 2021, The contribution of N-rich stars to the Galactic stellar halo using APOGEE red giants, MNRAS, 500, 5462

Kayser, A., Hilker, M., Grebel, E. K., & Willemsen, P. G. 2008, Comparing CN and CH line strengths in a homogeneous spectroscopic sample of 8 Galactic globular clusters, A&A, 486, 437

Koch, A., Grebel, E. K., & Martell, S. L. 2019, Purveyors of fine halos: Re-assessing globular cluster contributions to the Milky Way halo buildup with SDSS-IV, A&A, 625, A75

Koppelman, H. H., Helmi, A., Massari, D., Price-Whelan, A. M., & Starkenburg, T. K. 2019, Multiple retrograde substructures in the Galactic halo: A shattered view of Galactic history, A&A, 631, L9

Kupka, F., Piskunov, N., Ryabchikova, T. A., Stempels, H. C., & Weiss, W. W. 1999, VALD−2: Progress of the Vienna Atomic Line Data Base, A&AS, 138, 119

Lardo, C., Milone, A. P., Marino, A. F., et al. 2012, C and N abundances of main sequence and subgiant branch stars in NGC 1851, A&A, 541, A141

Lardo, C., Battaglia, G., Pancino, E., et al. 2016, Carbon and nitrogen abundances of individual stars in the Sculptor dwarf spheroidal galaxy, A&A, 585, A70

Lee, Y. S., Beers, T. C., Sivarani, T., et al. 2008a, The SEGUE Stellar Parameter Pipeline. I. Description and Comparison of Individual Methods, AJ, 136, 2022

Lee, Y. S., Beers, T. C., Sivarani, T., et al. 2008b, The SEGUE Stellar Parameter Pipeline. II. Validation with Galactic Globular and Open Clusters, AJ, 136, 2050

Lee, Y. S., Beers, T. C., Allende Prieto, C., et al. 2011, The SEGUE Stellar Parameter Pipeline. V. Estimation of Alpha-element Abundance Ratios from Low-resolution SDSS/SEGUE Stellar Spectra, AJ, 141, 90

Lee, Y. S., Beers, T. C., Masseron, T., et al. 2013, Carbon-enhanced Metal-poor Stars in SDSS/SEGUE. I. Carbon Abundance Estimation and Frequency of CEMP Stars, AJ, 146, 132

Lee, Y. S., Beers, T. C., Carlin, J. L., et al. 2015, Application of the SEGUE Stellar Parameter Pipeline to LAMOST Stellar Spectra, AJ, 150, 187

Luo, A.-L., Zhao, Y.-H., Zhao, G., & et al. 2019, VizieR Online Data Catalog: LAMOST DR5 catalogs (Luo+, 2019), VizieR Online Data Catalog, V/164

Majewski, S. R., Schiavon, R. P., Frinchaboy, P. M., et al. 2017, The Apache Point Observatory Galactic Evolution Experiment (APOGEE), AJ, 154, 94

Markwardt, C. B. 2009, Non-linear Least-squares Fitting in IDL with MPFIT, Astronomical Data Analysis Software and Systems XVIII, 411, 251

Martell, S. L., & Grebel, E. K. 2010, Light-element abundance variations in the Milky Way halo, A&A, 519, A14

Martell, S. L., Smolinski, J. P., Beers, T. C., & Grebel, E. K. 2011, Building the Galactic halo from globular clusters: evidence from chemically unusual red giants, A&A, 534, A136

Martell, S. L., Shetrone, M. D., Lucatello, S., et al. 2016, Chemical Tagging in the SDSS-III/APOGEE Survey: New Identifications of Halo Stars with Globular Cluster Origins, ApJ, 825, 146

Masseron, T., Plez, B., Primas, F., Van Eck, S., & Jorissen, A. 2006, A VLT-UVES spectrscopic analysis of C-rich Fe-poor stars, PhD thesis, Observatoire de Paris, France

Masseron, T., Plez, B., Van Eck, S., et al. 2014, CH in stellar atmospheres: an extensive linelist, A&A, 571, A47

Martin, Nicolas F., Venn, Kim A., Aguado, David S., et al. 2022, A stellar stream remnant of a globular cluster below the metallicity floor, NATURE, 601, 45

Myeong, G. C., Evans, N. W., Belokurov, V., Sanders, J. L., & Koposov, S. E. 2018, Discovery of new retrograde substructures: the shards of $\omega$ Centauri ?, MNRAS, 478, 5449

Myeong, G. C., Vasiliev, E., Iorio, G., Evans, N. W., & Belokurov, V. 2019, Evidence for two early accretion events that built the Milky Way stellar halo, MNRAS, 488, 1235

Naidu, R. P., Conroy, C., Bonaca, A., et al. 2020, Evidence from the H3 Survey That the Stellar Halo Is Entirely Comprised of Substructure, ApJ, 901, 48

Plez, B. 2012, Turbospectrum: Code for spectral synthesis, Astrophysics Source Code Library, ascl:1205.004

Re Fiorentin, P., Bailer-Jones, C. A. L., Lee, Y. S., et al. 2007, Estimation of stellar atmospheric parameters from SDSS/SEGUE spectra, A&A, 467, 1373

Schiavon, R. P., Zamora, O., Carrera, R., et al. 2017, Chemical tagging with APOGEE: discovery of a large population of N-rich stars in the inner Galaxy, MNRAS, 465, 501

Smith, V. V., Cunha, K., Shetrone, M. D., et al. 2013, Chemical Abundances in Field Red Giants from High-resolution H-band Spectra Using the APOGEE Spectral Linelist, ApJ, 765, 16

Smolinski, J. P., Lee, Y. S., Beers, T. C., et al. 2011, The SEGUE Stellar Parameter Pipeline. IV. Validation with an Extended Sample of Galactic Globular and Open Clusters, AJ, 141, 89

Smolinski, J. P., Martell, S. L., Beers, T. C., & Lee, Y. S. 2011b, A Survey of CN and CH Variations in Galactic Globular Clusters from Sloan Digital Sky Survey Spectroscopy, AJ, 142, 126

Suda, T., Katsuta, Y., Yamada, S., et al. 2008, Stellar Abundances for the Galactic Archeology (SAGA) Database — Compilation of the Characteristics of Known Extremely Metal-Poor Stars, PASJ, 60, 1159

Tang, B., Fernández-Trincado, J. G., Liu, C., et al. 2020, On the Chemical and Kinematic Consistency between N-

rich Metal-poor Field Stars and Enriched Populations in Globular Clusters, ApJ, 891, 28

Yanny, B., Rockosi, C., Newberg, H. J., et al. 2009, SEGUE: A Spectroscopic Survey of 240,000 Stars with g = 14-20, AJ, 137, 4377

York, D. G., Adelman, J., Anderson, J. E., et al. 2000, The Sloan Digital Sky Survey: Technical Summary, AJ, 120, 1579